\documentstyle[12pt]{article}
\begin{document}
\begin{center}
{\Large\bf Causal Propagators for the Second Order Wilson Loop} 
\vspace{1.0cm}
\\
B. M. Pimentel\footnote{Partially supported by CNPq}
and J. L. Tomazelli\footnote{Supported by CAPES}
\vspace{0.2cm}

Instituto de F\'{\i}sica Te\'{o}rica \\
Universidade Estadual Paulista \\
Rua Pamplona, 145 \\
01405--900 - S\~{a}o Paulo, SP - Brazil

\vspace{1.0cm}

{\bf Abstract}
\end{center}
\vspace{0.3cm}
 
We evaluate the Wilson loop at second order in general non-covariant gauges
by means of the causal principal-value prescription for the gauge-dependent
poles in the gauge-boson propagator and show that the result agrees with the
usual causal prescriptions.  
    
\vspace{0.4cm}

\vspace{1.0cm}

Its well known that the QCD vacuum structure is particularly simple in the
light-front formalism; for instance, the infinity-momentum frame, which fits
in this formalism, is appropriate to the study of deeply-inelastic scattering
processes. The rich underlying phenomenology has raised a lot of interest in
the study of non-Abelian gauge theories in general axial-type gauges, despite
some mathematical shortcomings that, fortunately, can be controled by
physical requirements, where causality plays the most important role.   

In the context of the gauge-invariant dynamics of QCD, the Wilson loop is 
certainly one of the most relevant objects in such theories, from the
conceptual standpoint. In 1982, the computation of the Wilson loop to the
fourth order carried out by Caracciolo {\it et al\/}${}^{[1]}$ has revealed
that in the temporal gauge the na\"{\i}ve Cauchy principal-value (PV)
prescription used to handle the gauge-dependent poles in the gauge-boson
propagator leads to results which fails to agree with the ones obtained in
the Feynman and Coulomb gauges. Later, Basseto {\it et al\/}${}^{[2]}$
arrived at the correct result using the unified-gauge formalism introduced by
Leibbrandt${}^{[3]}$.  

In 1991, Pimentel and Suzuki${}^{[4]}$ have proposed a causal principal-value
prescription for the light-cone gauge by conjecturing that the gauge-boson 
propagator as a whole must be causal${}^{[5]}$. Recently causal propagators
for non-covariant gauges were derived on the same grounds as for covariant
radial distributions${}^{[6]}$ and shown to coincide with the resulting 
distributions obtained through the causal prescription of Pimentel and
Suzuki, except for the pure axial gauge${}^{[7]}$. 

In the present work we test the consistency of the Pimentel-Suzuki
prescription by calculating the Wilson loop at one-loop order following the
ma\-nifestly gauge-invariant procedure of Hand and Leibbrandt${}^{[8]}$, who
have used distinct sets of vectors ${n_{\mu},n_{\mu}^*}$ and 
${N_{\mu},N_{\mu}^*}$ for the paths and gauge-fixing constraint,
respectively.    

The Lagrangian density for the massless Yang-Mills theory is given by
\begin{equation}
{\cal L}=-\frac{1}{4}F_{\mu\nu}^aF^{\mu\nu}_a-\frac{1}{2\alpha}
(N.A^a)(N.A_a)\,\,,\,\,\alpha\rightarrow 0\,\,,
\end{equation}
where $N_{\mu}=(N_0,{\bf N})$ is the gauge-fixing vector, and
\begin{equation}
N^{\mu}A_{\mu}^a=0\,\,,\,\,\mu=0,\dots 3\,\,,
\end{equation}
the gauge-fixing constraint.

The one-loop expectation value of the Wilson loop for a rectangular path
lying in Minkowski space, characterized in terms of two light-cone vectors
$n_{\mu}=(n_0,{\bf n})$ and $n_{\mu}^*=(n_0,-{\bf n})$, is 
\begin{eqnarray}
W^{(1)}=(ig^2)C_F\mu^{4-D}\int\frac{d^Dk}{(2\pi)^D}G_{\mu\nu}(k)\int_0^1dt
\int_0^1dt'[n_{\mu}^*n_{\nu}^*F_1(t,t') \nonumber\\
+n_{\mu}n_{\nu}F_2(t,t')+n_{\mu}n_{\nu}^*F_3(t,t')]\,\,,
\end{eqnarray}
where
\begin{eqnarray}
F_1(t,t') \equiv e^{ik.n^*(t-t')}-e^{-ik.n^*(t-t')+ik.n}\,\,,\\
F_2(t,t') \equiv e^{ik.n(t-t')}-e^{-ik.n(t-t')-ik.n^*}\,\,,\\
F_3(t,t') \equiv e^{-ik.n^*t+ik.nt'+ik.n^*}-e^{-ik.nt+ik.n^*t'+ik.(n-n^*)}
\nonumber\\
+e^{-ik.n^*t+ik.nt'-ik.n}-e^{ik.n^*t-ik.nt'}\,\,,
\end{eqnarray}
and
\begin{equation}
G_{\mu\nu}(k)=\frac{-i}{k^2+i\epsilon}\left[g_{\mu\nu}-\frac{k_{\mu}N_{\nu}+
k_{\nu}N_{\mu}}{k.N}+\frac{N^2}{(k.N)^2}k_{\mu}k_{\nu}\right]
\end{equation}
is the gluon propagator.
 
Performing the integration over the path variables $t$ and $t'$ we obtain
after contraction of the Lorentz indices 
\begin{eqnarray}
W^{(1)}=(ig)^2C_F\mu^{4-D}2n.n^*\int\frac{d^Dk}{(2\pi)^D}\frac{-i}
{k^2+i\epsilon}\frac{1}{(k.n)(k.n^*)} \nonumber\\
\times[-2+2e^{ik.n}+2e^{ik.n^*}-e^{ik.(n+n^*)}-e^{ik.(n-n^*)}]\,\,.
\end{eqnarray}

Instead of choosing any {\it ad hoc} prescription to treat the poles 
$(k.N)^{-1}$ and $(k.N)^{-2}$ in (7) we now apply the principle of analytic
continuation in order to derive the causal distribution corresponding to the
gauge-boson propagator. For this purpose let us consider the product
$[k^2(k.n)^m]^{-1}$ with $n^{\mu}\equiv (n^0,0,0,n^3)$ being an external
arbitrary vector.The factor $[k^2(k.n)^m]^{-1}$ upon the hypothesis of
analytic continuation to the upper complex half-plane becomes
\begin{equation}
\frac {1}{k^2(k.n)^m} \rightarrow \frac {1}{(k^2+2i\epsilon
k_0^2)(k.n + i\epsilon k^0n^0)^m} .
\end{equation}

Due to the arbitrariness of $n$, it can be chosen so that $n^0>0$, and since 
$\epsilon $ is strictly positive, equation (9) becomes   
\begin{eqnarray}
\frac {1}{k^2(k.n)^m} & \rightarrow & \frac {1}{(k^2+2i\epsilon
k_0^2)(k.n+i\epsilon|k^0|n^0)^m},\hspace{0.5cm} \mbox{for}
\hspace{0.1cm} k^0>0 \nonumber \\
\frac {1}{k^2(k.n)^m} & \rightarrow & \frac
{1}{(k^2+2i\epsilon k_0^2)(k.n-i\epsilon|k^0|n^0)^m},
\hspace{0.5cm} \mbox{for} \hspace{0.1cm} k^0<0  
\end{eqnarray}
or, using the Heaviside distribution,
\begin{equation}
\frac{1}{k^2(k.n)^m} \rightarrow \frac{1}{k^2+i\varepsilon
}\left\{\frac {\Theta(-k^0)}{(k.n-i\xi)^m}+\frac{\Theta(k^0)}
{(k.n+i\xi)^m}\right\},\,\,
\begin{array}{c}\varepsilon  
\equiv2\epsilon k_0^2 \rightarrow 0^{+} \\
\xi \equiv \epsilon|k^0|n^0 \rightarrow 0^{+} \end{array},
\end{equation}
which is just the causal prescription considered in reference [2] for m=2.
We can extend the above derivation to the case $n^0=0$ if, before analytic
continuing $k^0$, we first perform an infinitesimal Lorentz transformation
\begin{equation}
n^{\mu} \rightarrow n'^{\mu}=\Lambda^{\mu}{}_{\nu}\,n^{\nu}=
\Lambda^{\mu}{}_3\,n^3\,\,,\,\,\,\Lambda^0{}_3\,n^3>0\,\,,
\end{equation}
and then return to the original Lorentz frame, so that the gauge condition
(2) is preserved. 

Thus, making the substitution (11) in equation (8) for m=1 in order to treat
the pole $(k.n)^{-1}$, and using the distribution identity
\begin{equation}
\frac{1}{k.n \pm i\xi}=PV\frac{1}{k.n} \mp i\pi\delta(k.n)\,\,,
\end{equation}
we arrive at 
\begin{equation}
W^{(1)}=(ig^2)C_F\mu^{4-D}2n.n^*(I_{PV}+I_{\delta})\,\,,
\end{equation}
where we have defined
\begin{eqnarray}
I_{PV}\equiv\frac{1}{2}\int\frac{d^Dk}{(2\pi)^D}\frac{-i}{k^2+i\epsilon}
\left\{\frac{1}{k.n+i\xi}+\frac{1}{k.n-i\xi}\right\}\frac{1}{k.n^*}
\nonumber\\
\times[-2+2e^{ik.n}+2e^{ik.n^*}-e^{ik.(n+n^*)}-e^{ik.(n-n^*)}] \nonumber\\
=\frac{1}{2}\int\frac{d^Dk}{(2\pi)^D}\frac{-i}{k^2+i\epsilon}
\left\{\frac{1}{(k.n)(k.n^*)+i\xi}+\frac{1}{(k.n)(k.n^*)-i\xi}\right\}
\nonumber\\
\times[-2+2e^{ik.n}+2e^{ik.n^*}-e^{ik.(n+n^*)}-e^{ik.(n-n^*)}]\,\,,\\
I_{\delta}\equiv -i\pi\int\frac{d^Dk}{(2\pi)^D}\frac{-i}{k^2+i\epsilon}
\frac{1}{k.n^*}\varepsilon(k^0)\delta(k.n) \nonumber\\
\times[-2+2e^{ik.n}+2e^{ik.n^*}-e^{ik.(n+n^*)}-e^{ik.(n-n^*)}]\,\,.
\end{eqnarray}

Performing the integral over $k_0$, we see that $I_{\delta}$ vanishes by 
symmetric integration:
\begin{equation}
I_{\delta}=-i\pi n_0\int\frac{d^{D-1}{\bf k}}{(2\pi)^D}
\frac{1}{[({\bf k}.{\bf n})^2-n_0^2{\bf k}^2]}\frac{\sin(2{\bf k}.{\bf n})} 
{|{\bf k}.{\bf n}|}=0\,\,,
\end{equation}
since the integrand is an odd function in the components of the vector 
${\bf k}$. On the other hand, making use of identity (13) once more, we
rewrite the remaining integral $I_{PV}$ in the form
\begin{eqnarray}
I_{PV}=\int\frac{d^Dk}{(2\pi)^D}\frac{-i}{k^2+i\epsilon}
\frac{1}{(k.n)(k.n^*)+i\xi}\nonumber\\
\times[-2+2e^{ik.n}+2e^{ik.n^*}-e^{ik.(n+n^*)}-e^{ik.(n-n^*)}]+I_R\,\,,
\end{eqnarray}
where
\begin{eqnarray}
I_R= i\pi\int\frac{d^Dk}{(2\pi)^D}\frac{-i}{k^2+i\epsilon}
\delta[(k_0.n_0)^2-({\bf k}.{\bf n})^2] \nonumber\\
\times[-2+2e^{ik.n}+2e^{ik.n^*}-e^{ik.(n+n^*)}-e^{ik.(n-n^*)}]\,\,,
\end{eqnarray}
which also vanishes by symmetric integration.

Consequently, we obtain for the Wilson loop (14) exactly the same expression
as the corresponding in reference [8]:
\begin{eqnarray}
W^{(1)}=(ig^2)C_F\mu^{4-D}2n.n^*\int\frac{d^Dk}{(2\pi)^D}
\frac{-i}{k^2+i\epsilon}\frac{1}{(k.n)(k.n^*)+i\xi} \nonumber\\
\times[-2+2e^{ik.n}+2e^{ik.n^*}-e^{ik.(n+n^*)}-e^{ik.(n-n^*)}] \nonumber\\
=(ig^2)C_F\mu^{4-D}\frac{4\pi^{D/2}}{(2\pi)^D}i^{2-D/2}\int_0^1dy\,
(1-y)^{1-D/2}\nonumber\\
\times\int_0^{\infty}dx\frac{e^{-x[(1-y)\epsilon+y\xi/n_0^2]}}{x^{D/2-1}}
\left[2-\exp(\frac{-in_0^2}{x})-\exp(\frac{i{\bf n}^2}{x})\right]\,\,.
\end{eqnarray}
Evaluation of (20) yields
\begin{eqnarray}
W^{(1)}=\frac{g^2C_F\mu^{4-D}}{(2\pi)^{D/2}}
\frac{4\Gamma(D/2-1)}{(4-D)^2}[(n_0^2+i\eta)^{2-D/2}\nonumber\\
+(-n_0^2+i\eta)^{2-D/2}-2(i\eta)^{2-D/2}] 
\,\,,\,\,\eta\rightarrow 0^{+}\,\,.
\end{eqnarray}
The last term in square brackets in the above equation is absent in the 
corresponding expression of reference [8]. However, this is of no
significance since we are considering the analytic extension of $W^{(1)}$ in
the strip \\
$3<Re\,D<4$ and, therefore, may set $\eta$ equal to zero before
making the expansion around $D=4-\epsilon$.    

From the quoted results we may conclude that the unified-gauge formalism and
the causal principal-value prescription are equivalent approaches to the
Wilson loop in second order of perturbation theory for general axial-type 
gauges. A similar result was found regarding the Mandelstam-Leibbrandt (ML)
and the ordinary PV prescription${}^{[9]}$ for particular gauge choices. 

\newpage

{\large\bf 6.\,References}
\vspace{1.0cm}

\begin{sf}

\begin{description}

\item[{[1]}] S. Caracciolo, G. Curci and P. Menotti, {\it Phys.Lett.} {\bf
113 B} (1982) 311;

\item[{[2]}] A. Basseto, I. A. Korchemskaya, G. P. Korchemsky and G.
Nardelli, {\it Nucl. Phys.} {\bf B 408} (1993) 62;

\item[{[3]}] G. Leibbrandt, {\it Nucl. Phys.} {\bf B 310} (1988) 405;

\item[{[4]}] B. M. Pimentel and A. T. Suzuki, {\it Mod. Phys.Lett. A} {\bf
6} (1991) 2649;

\item[{[5]}] B. M. Pimentel and A. T. Suzuki, {\it Phys. Rev.} {\bf D 42}
(1990) 2115;

\item[{[6]}] C. G. Bollini and J. J. Giambiagi, {\it Il Nuovo Cimento} {\bf
34} (1965) 1146;

\item[{[7]}] B. M. Pimentel, A. T. Suzuki and J. L. Tomazelli, preprint 
hep-th/9510196, to appear in {\it Il Nuovo Cimento} {\bf B} (1996);

\item[{[8]}] B. J. Hand and G. Leibbrandt, preprint CERN-TH/95-227;

\item[{[9]}] H. H\"uffel, P. V. Landshoff and J. C. Taylor, {\it Phys. Lett.}
{\bf B 217} (1989) 147.
\end{description}

\end{sf}
\end{document}